\documentclass{emulateapj}
\usepackage{pstricks,apjfonts}

\def\asca       {{\em ASCA}\/}
\def\rosat      {{\em ROSAT}\/}
\def\chandra    {{\em Chandra}\/}
\def\xmm        {{\em XMM}\/}

\def\as         {$^{\prime\prime}$}
\def\cmsq       {cm$^{-2}$}
\def\kmsmpc     {km$\;$s$^{-1}\,$Mpc$^{-1}$}
\def\gax        {\gtrsim}
\def\lax        {\lesssim}

\begin{document}

\submitted{ApJ Letters, in press; astro-ph/0301367} 

\lefthead{TEMPERATURE MAP OF A754 AND THERMAL CONDUCTION}
\righthead{MARKEVITCH ET AL.}

\title{{\em CHANDRA}\/ TEMPERATURE MAP OF ABELL 754 AND CONSTRAINTS ON
THERMAL CONDUCTION}

\author{M.~Markevitch, P.~Mazzotta$^1$, A.~Vikhlinin$^2$, D.~Burke, Y.~Butt,
L.~David, H.~Donnelly, W.~R.~Forman, D.~E.~Harris, D.-W.~Kim, S.~Virani,
J.~Vrtilek}

\affil{Harvard-Smithsonian Center for Astrophysics, 60 Garden St.,
Cambridge, MA 02138; maxim@head-cfa.harvard.edu}

\altaffiltext{1}{University of Durham, U.K.}
\altaffiltext{2}{Also IKI, Moscow, Russia}

\begin{abstract}

We use \chandra\ data to derive a detailed gas temperature map of the
nearby, hot, merging galaxy cluster A754.  Combined with the X-ray and
optical images, the map reveals a more complex merger geometry than
previously thought, possibly involving more than two subclusters or a cool
gas cloud sloshing independently from its former host subcluster.  In the
cluster central region, we detect spatial variations of the gas temperature
on all linear scales, from 100 kpc (the map resolution) and up, which likely
remain from a merger shock passage. These variations are used to derive an
upper limit on effective thermal conductivity on a 100 kpc scale, which is
at least an order of magnitude lower than the Spitzer value.  This
constraint pertains to the bulk of the intracluster gas, as compared to the
previously reported estimates for cold fronts (which are rather peculiar
sites).  If the conductivity in a tangled magnetic field is at the recently
predicted higher values (i.e., about 1/5 of the Spitzer value), the observed
suppression can be achieved, for example, if the intracluster gas consists
of magnetically isolated domains.

\end{abstract}


\section{INTRODUCTION}

Since the non-detection by \xmm\ and \chandra\ of the expected quantities of
cool gas in cluster central regions (Peterson et al.\ 2001 and later works),
there has been a renewed interest in heat conduction as a possible mechanism
to off-set radiative cooling (e.g., Narayan \& Medvedev 2001; Gruzinov 2002;
Fabian, Voigt, \& Morris 2002).  Detailed analyses of several cooling flow
clusters (Voigt et al.\ 2002; Zakamska \& Narayan 2003) show that heat
inflow from the outer regions may indeed be sufficient over most of the
cooling region, but only if the conductivity, $\kappa$, is not lower than
$\kappa_S/3$, where $\kappa_S$ is the Spitzer (1962) value.  Such high
conduction would have other important effects, e.g., erasing the
cluster-scale temperature gradients (e.g., David, Hughes, \& Tucker 1992)
and causing significant heat loss into the intercluster space (Loeb 2002).

The conductivity of the cluster plasma should be reduced by the magnetic
fields; however, theoretical estimates for such a reduction have not yet
converged (see \S\ref{sec:cond} for details).  On the observational side,
David et al.\ (1992) derived a lower limit of $\kappa > \kappa_S/10$ for the
Coma cluster.  However, recent higher-resolution data revealed temperature
variations in the cluster region that those authors assumed isothermal
(e.g., Neumann et al.\ 2002), which may weaken their constraint.  The only
cluster-scale results derived so far using the high-resolution data were for
the conduction across cold fronts.  Ettori \& Fabian (2000) pointed out that
the temperature jumps across the fronts in A2142 observed by \chandra\
(Markevitch et al.\ 2000) imply $\kappa < \kappa_S/250$.  Although one can
argue that as the cold front moves, the temperature gradient may be
continually sharpened because of stripping of the heated boundary layer,
their conclusion should be correct.  As shown by Vikhlinin, Markevitch \&
Murray (2001c) for A3667, there is no such boundary layer and the gas
density discontinuity at the front is very sharp --- much narrower than the
electron mean free path, indicating that diffusion and conduction across the
front are indeed suppressed. The likely reason is a layer of the magnetic
field parallel to the front, which can form as the gas flows around the
front and straightens the tangled field lines (Vikhlinin, Markevitch \&
Murray 2001b).  Another place where conductivity was estimated is the
boundary between the hot cluster gas and two cool, galaxy-size dense gas
clouds in Coma (Vikhlinin et al.\ 2001a).  For these clouds to survive,
conduction across their boundaries should be reduced by a factor 30--100.

Those measurements probe rather peculiar sites in clusters, i.e.\ boundaries
between different gas phases which are expected to have disjoint magnetic
fields and so be thermally isolated.  As we will show below, an estimate
that is more representative of the bulk of the gas can be obtained from the
temperature maps of hot merging clusters, such as A754.

Previous optical and X-ray studies of A754 --- a rich cluster at $z=0.054$
--- showed that it is the prototype of a major merger.  It has a complex
galaxy distribution, X-ray morphology and gas temperature structure
(Fabricant et al.\ 1986; Bird 1994; Slezak, Durret, \& Gerbal 1994;
Zabludoff \& Zaritsky 1995; Henry \& Briel 1995; Henriksen \& Markevitch
1996, hereafter HM96).  It also exhibits a radio halo (Kassim et al.\ 2001).
In this paper, we use \chandra\ data to derive a detailed gas temperature
map for A754.  In \S\ref{sec:merg}, we discuss the merger details revealed
by the new data.  In \S\ref{sec:cond}, our temperature map is used for an
estimate of thermal conductivity of the cluster gas. At this redshift,
$1''=1.13$ kpc for our assumed $H_0=65\;h_{65}$ \kmsmpc.

\begin{figure*}[t]
\pspicture(0,14.0)(18.5,23.8)

\rput[tl]{0}(0.95,24){\epsfysize=10.6cm \epsfclipon
\epsffile{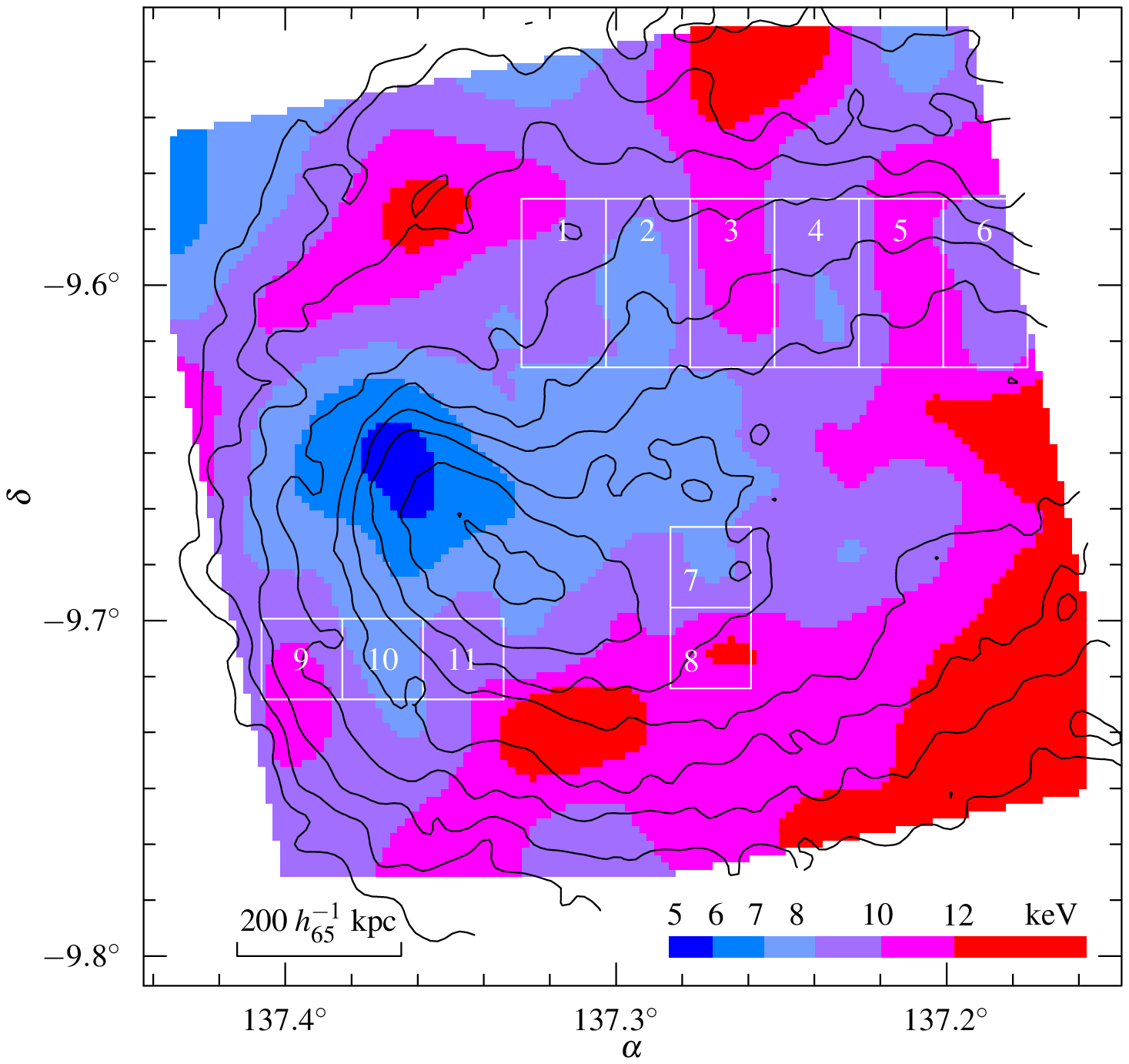}}

\rput[tl]{0}(11.35,24){\epsfysize=10.6cm \epsfclipon
\epsffile{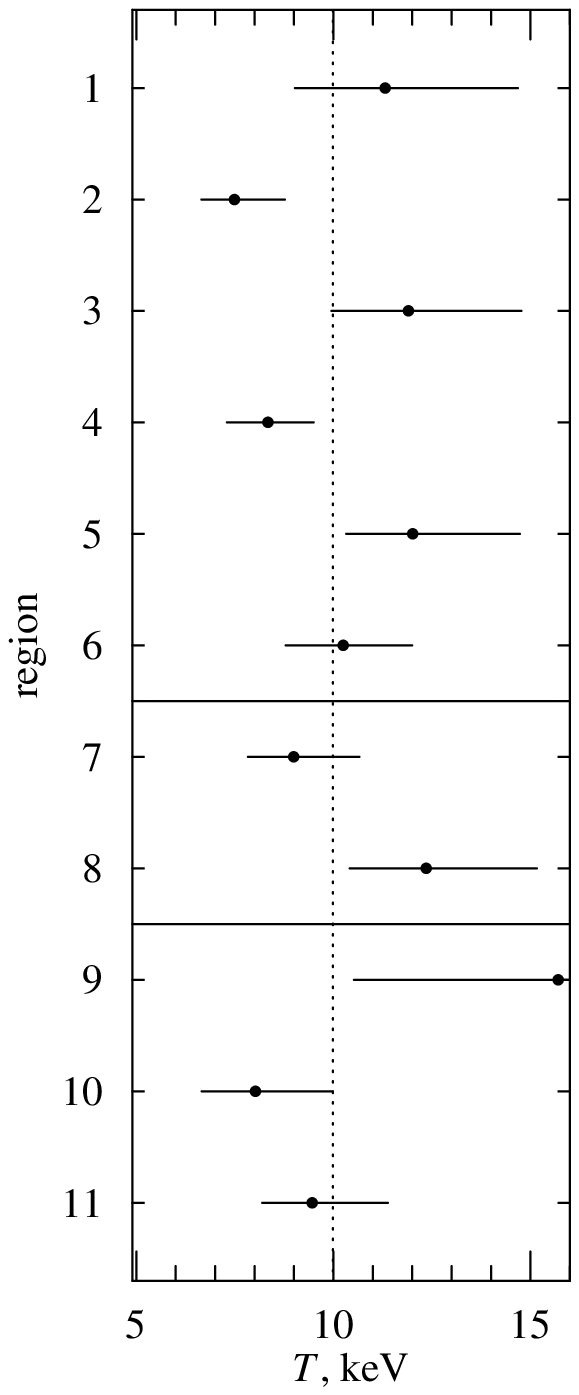}}

\endpspicture

\caption{Projected temperature map of A754 (colors) overlaid on the ACIS
  0.8--5 keV image (contours, spaced by $\sqrt 2$). Point sources are
  excluded.  The map is adaptively smoothed (as part of the fitting
  procedure) by a Gaussian with $\sigma$ between 25\as\ and 55\as; the image
  is smoothed with $\sigma$ between 10\as\ and 14\as. In most areas,
  different colors correspond to significantly different temperatures.
  Direct temperature fits for regions shown by white rectangles are given in
  the right panel (errors are 90\%). The vertical dotted line shows the
  cluster average temperature. The step of the rectangles is 100 kpc.}
\label{fig:tmap}
\end{figure*}

\section{DATA ANALYSIS} 

A754 was observed by \chandra\ in October 1999 using ACIS-I%
\footnote{\chandra\ Observatory Guide
http://asc.harvard.edu/udocs/docs/, ``Observatory Guide,'' ``ACIS''}
for a clean exposure of 39 ks.  Our analysis procedure, briefly outlined
below, follows that in Markevitch et al.\ (2000) and Markevitch \& Vikhlinin
(2001) with a few updates.  For background subtraction, we used the ACIS
blank-sky dataset normalized by the ratio of the 10--12 keV rates in the
data and the background files (see Markevitch et al.\ 2003); this
normalization was within 1\% of their exposure ratio.  From the \rosat\
All-Sky Survey (Snowden et al.\ 1997), there is no soft Galactic excess at
the cluster position so no additional background modeling was necessary.
The detector low energy response was corrected for the secular change of the
ACIS quantum efficiency (Plucinsky et al.\ 2002).  All point sources were
masked out.

A single-temperature fit to the spectrum for the whole cluster (as far as we
could reach, $r=9'=0.6\;h_{65}^{-1}$ Mpc) gives $T_e=10.0\pm 0.3$ keV and a
metal abundance of $0.30\pm 0.05$ (relative to Anders \& Grevesse 1989) at
the 90\% confidence.  This temperature is somewhat higher than the \asca\
value of $8.5-9.0$ keV (HM96); however, \asca\ covered more of this highly
nonisothermal cluster.  The absorption column, if fitted as a free
parameter, is in good agreement with the Galactic value of $N_H=4.36\times
10^{20}$ \cmsq, and therefore was fixed.

To derive a temperature map, we extracted cluster images in 7 energy bands
between $0.8-9.0$ keV, excluding the intervals $1.8-2.2$ keV because of the
poor calibration and $7.4-7.6$ keV containing a bright fluorescent
background line.  These images were adaptively smoothed (identically in all
energy bands).  In each image pixel, the 7 flux values were fitted by a
thermal model with fixed $N_H$ and abundances.  The result is shown in
Fig.~\ref{fig:tmap}.

\section{DISCUSSION}

The large-scale temperature structure is in broad agreement with earlier
\rosat\ and \asca\ results (Henry \& Briel 1995; HM96).  The new map offers
a much more detailed look into this interesting cluster.  Of note are a
large hot area to the south and southwest (already seen in the earlier
data), cool gas at the northeast tip of the elongated central bright body,
and numerous small-scale temperature variations.  Most of these variations
are statistically significant; to illustrate this, we extracted and fitted
the real spectra for several less-convincing features (right panel in
Fig.~\ref{fig:tmap}).  There is also a curious wavy structure in both the
X-ray brightness and the temperature map (in the general area of regions
1--6 in Fig.~\ref{fig:tmap}), whose origin we find hard to explain.

\begin{figure}[t]
\pspicture(0,15.9)(18.5,24.3)

\rput[tl]{0}(0.4,24){\epsfxsize=8.0cm \epsfclipon
\epsffile{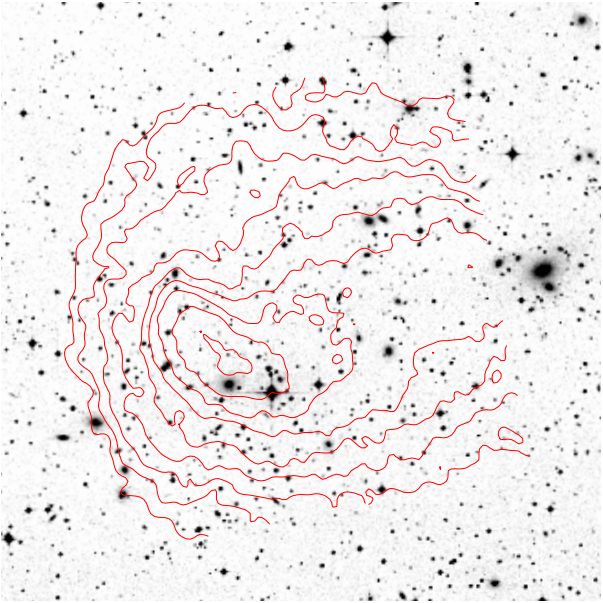}}

\endpspicture

\caption{\chandra\ X-ray brightness contours (same as in Fig.~1) overlaid on
  the Palomar Digitized Sky Survey optical image.}
\label{fig:img}

\end{figure}

\subsection{Merger geometry}
\label{sec:merg}

HM96 proposed that A754 is an off-center merger, and Roettiger, Stone, \&
Mushotzky (1998) succeeded in reproducing the general large-scale X-ray
structure in hydrodynamic simulations.  In Fig.\ 2, which overlays X-ray and
optical images, one can discern two main galaxy concentrations, one around
the western brightest galaxy, another just south of the X-ray brightness
peak (e.g., Fabricant et al.\ 1986; Zabludoff \& Zaritsky 1995).  While the
western X-ray extension (not completely covered by our data, but see, e.g.,
Henry \& Briel 1995) is associated with the western galaxy concentration,
the second galaxy subcluster is offset from its presumed X-ray counterpart,
the bright elongated core (Zabludoff \& Zaritsky 1995).  In the Roettiger et
al.\ simulations, this elongation and offset are caused by the ram pressure
exerted from the southeast by the gas that used to belong to the western
galaxy subcluster (which is now moving further to the northwest).

However, the \chandra\ data complicate this picture by revealing interesting
structure within the elongated gas core.  There is a steep X-ray brightness
drop-off at the northern tip of the body (Figs.\ 1, 2) which, although not as
sharp and regular-shaped as some cold fronts (e.g., Vikhlinin et al.\
2001b), indicates a similar moving interface.  The elongated cloud
apparently moves to the north (in projection) from the big galaxy
concentration, not being swept away by ram pressure from the south, but as
if dragged along by another subcluster.  The cold region at the nose of the
elongated body appears to be the former core of that subcluster.  However,
it does not coincide with any large galaxy concentration (although there is
a possible small group there, which alone would not be considered
statistically significant).  Alternatively, in the course of the merger, the
elongated cloud may have completely decoupled from its former dark matter
host and is now ``sloshing'' independently.  Such a state would be very
short-lived, because a moving gas cloud without the dark matter gravity
support is quickly destroyed by gas-dynamic instabilities.

The coolest spot does not coincide with the gas density peak, so apparently
it is not in pressure equilibrium with the rest of the elongated core.  Such
a transient state may arise, for example, if the cool core travels down the
strong ambient pressure gradient faster than its internal sound speed.  Its
northern tip may then expand too fast for the internal pressure to equalize.
It is also possible that projection effects are at play (e.g., the elongated
core is a projection of two clouds).

We finally mention a brightness feature that looks like a bow shock leaving
the merger site, first noted in the \rosat\ image (Krivonos et al.\ 2003).
It is located at the leftmost brightness contour in our figures
(unfortunately, we cannot study it because of the limited coverage).  It is
consistent with the general merger picture proposed by Roettiger et al.

To conclude, simple scenarios cannot explain all the complex details in the
new X-ray data on A754.  However, the very complexity of the temperature map
of this cluster can be used for an interesting physical estimate, as we show
below.

\subsection{Limit on thermal conduction}
\label{sec:cond}

In a plasma with a temperature gradient $\nabla T_e$ such that the
gradient's linear scale $l_T \equiv T_e/|\nabla T_e| \gg \lambda_e$ (where
$\lambda_e$ is electron mean free path), heat flux is given by $q=-\kappa
\nabla T_e$.  In the absence of magnetic fields and for typical cluster
plasma parameters, the collisional conductivity is given by Spitzer (1962):
\begin{eqnarray}
\label{eq:kappa}
\kappa_S &\; =\; & 1.31\, n_e\, \lambda_e\, k\; ( kT_e / m_e)^{1/2} \\
         &\;\simeq\; & 7.1\times 10^{13} 
         \left(\frac{kT_e}{10\;{\rm keV}}\right)^{5/2} \!
         \left(\frac{\ln \Lambda}{37.8}\right)^{-1}
            {\rm erg\;s}^{-1}{\rm cm}^{-1}{\rm K}^{-1},  \nonumber
\end{eqnarray}
where $\ln \Lambda$ is the Coulomb logarithm.  For magnetic fields expected
in clusters ($B>0.1 \mu G$, Carilli \& Taylor 2002), the electron and proton
gyro-radii are many orders of magnitude smaller than their Coulomb mean free
paths.  Therefore, conduction occurs predominantly along the magnetic field
lines, and the effective conductivity is determined by the topology of the
field.  If the field is chaotic with a coherence scale $l_B \gg \lambda_e$,
the conductivity on scales $l_T\gg l_B$ is reduced by a factor of 3 (e.g.,
Sarazin 1988; Gruzinov 2002).  If $l_B \lax \lambda_e$ (which is likely the
case --- radio data suggest $l_B \sim 5-30$ kpc, e.g., Carilli \&
Taylor 2002, while $\lambda_e\simeq 10-15$ kpc in the central region of A754),
the effective conductivity is reduced because electrons have to travel a
longer path along the tangled field lines to diffuse over a given distance.
The reduction is of order $(l_T/l_B)^2$ (e.g., Tribble 1989; note that
$\kappa$ becomes scale-dependent), but probably not more than a factor of
100 if one considers the physical origin of the tangled field (Rosner \&
Tucker 1989; Tao 1995).  In addition, the conductivity {\em along}\/ the
field lines may be reduced by a factor of $\sim 10$ (for $l_B \sim
\lambda_e$) by magnetic mirrors in the stochastic field (Malyshkin \&
Kulsrud 2001).  On the other hand, for certain spectra of the field
fluctuations, diffusion across the field lines and divergence of the lines
may boost the effective conductivity to within a factor of only $\sim 5$
below $\kappa_S$ (Narayan \& Medvedev 2001).  This requires a field tangled
on a range of scales down to $\sim l_B/100$.  We note that it is unclear
whether such a field would be consistent with the observations of the $\sim
10$ kpc coherent ``patches'' of polarization of the background radio sources
(Carilli \& Taylor 2002; Govoni 2001), if those measurements indeed probe
typical intracluster fields.  To summarize, the collisional thermal
conductivity on 100 kpc scales is expected to be reduced by a factor
$3-100$.

Our temperature map of A754 offers an interesting opportunity to put
observational constraints on $\kappa$.  All over the cluster, the
temperature varies on linear scales starting from 100 kpc (the map
resolution, limited by statistics) and up.  This scale is so small that the
temperature gradients should be erased very quickly if the conductivity is
at the Spitzer value (\ref{eq:kappa}).  An approximate expression for the
conduction timescale $t_{\rm cond}\equiv -(d \ln T_e/dt)^{-1}$ can be found
in, e.g., Sarazin (1988):
\begin{eqnarray}
\label{eq:tcond}
t_{\rm cond} &\; \sim\; & k\, n_e\, l_T^2\,/\, \kappa_S \;
                                 \simeq\; 1.2\times 10^7\; {\rm yr}\nonumber \\
             & \times & 
\frac{n_e}{2\times 10^{-3}\; {\rm cm}^{-3}}
\left( \frac{l_T}{100\; {\rm kpc}} \right)^2
\left( \frac{k T_e}{10\;{\rm keV}} \right)^{-5/2},
\end{eqnarray}
where $n_e$ is from Jones \& Forman (1999); $t_{\rm cond} \propto
h_{65}^{-3/2}$.  Note that $l_T$ as defined above and determined from the
map in Fig.\ 1 would be greater than 100 kpc.  However, we measure
temperatures in projection, with a large column of gas at the mean cluster
temperature on the line of sight, so the true temperature
variations should be significantly stronger.%
\footnote{The average true distance between regions of different temperature
in the map is also greater, but it is the size of each region, not the
distance between them, that determines $l_T$.}
We therefore chose to use the apparent scale of the variations in the map as
a realistic value of $l_T$ for this estimate; this, along with the strong
dependence of $\kappa_S$ on $T_e$, are our greatest sources of uncertainty.
We did not take into account mild saturation of the
classical heat flux (for our $\lambda_e$ and $l_T$ one expects $\kappa
\simeq 0.7\;\kappa_S$, cf.\ Cowie \& McKee 1977 and Sarazin 1988), which in
reality should not occur because of the tangled fields.  In any case, it
would not affect the conclusion.

At the same time, we do see these temperature variations everywhere in the
cluster, so they should have lived for at least as long as it takes for a
merger shock and/or a disturbing subcluster (that presumably have created
them via compression and gas-dynamic instabilities) to travel across the
central region of the cluster, $L\gax 800$ kpc (approximately the size of
our map).  The sound speed in this region is a safe estimate of that shock's
velocity regardless of its Mach number, since this region now contains the
post-shock gas.  This timescale is
\begin{equation}
t_{\rm age}\; \gax\; \frac{L}{c_s}\; \simeq\; 5\times 10^8\;
                     \frac{L}{800\;{\rm kpc}} 
                     \left(\frac{kT_e}{10\;{\rm keV}}\right)^{-1/2} {\rm yr};
\label{eq:tage}
\end{equation}
$t_{\rm age} \propto h_{65}^{-1}$. Comparison of eqs.\ (\ref{eq:tcond}) and
(\ref{eq:tage}) shows that the conductivity should be reduced by a factor
\begin{equation}
(\kappa/\kappa_S)^{-1}\; \sim\; t_{\rm age}/t_{\rm cond}\; \gax\; 
40\; h_{65}^{1/2},
\label{eq:limit}
\end{equation}
although we emphasize that, given all the uncertainties, this is only an
order-of-magnitude estimate.

The limit in eq.\ (\ref{eq:limit}) is inconsistent with the upper end of the
theoretical range for $\kappa$.  We note that the ``local'' conductivity
considered by theorists can still be higher, if, for example, the ICM
consists of thermally isolated domains with self-contained field structure.
Such domains of gas that belonged to different subclusters may survive if
mixing during the merger is not very efficient, as is indeed suggested by
the existence of cold fronts (Vikhlinin et al.\ 2001b; Narayan \& Medvedev
2001).

Our upper limit on conduction appears to exclude it as a mechanism to
off-set cooling in the cluster cores (where one needs $\kappa>\kappa_S/3$,
see refs.\ in \S1).  However, conduction in a denser, cooler core is likely
to operate in a different regime; for example, there one expects $\lambda_e
\ll l_B$ compared to $\lambda_e \sim l_B$ in the regions that we studied
above.  In addition, the central magnetic fields may have a different
topology because of the competing effects of the radial cooling inflow
(Bregman \& David 1988; Soker \& Sarazin 1990) and radial infall of the
subcluster pieces vs.\ the widespread gas ``sloshing'' (Markevitch,
Vikhlinin, \& Forman 2002).  This may result in a different reduction
factor.  Finally, we note that our limit in eq.\ (\ref{eq:limit}), as well
as the cooling flow conduction estimates, are relative to $\kappa_S$ which
itself is much lower in the cool cores.  Our {\em absolute}\/ upper limit on
conductivity is not necessarily lower than the {\em absolute}\/ conductivity
needed to compensate cooling.  Thus, even if the reduction factor for the
Spitzer conduction is the same everywhere in the cluster, we cannot exclude
a sufficiently high heat influx into the cooling core that is due to a
different physical process (and so has a different dependence on gas
parameters), such as turbulence, for example.

\section{SUMMARY}

We have derived a detailed temperature map of the prototypical merging
cluster A754.  Simple merger scenarios cannot reproduce the complex details
revealed by the new data; either A754 is a three-body merger, or the cool
dense gas found in the cluster's bright elongated core has decoupled from
its former subcluster host and is sloshing independently.  We use the
small-scale structure in the temperature map to derive the first constraint
on thermal conductivity in the bulk of the gas in a cluster.  The
conductivity on scales $\sim 100$ kpc appears reduced by at least an order
of magnitude from the Spitzer value.

\acknowledgements

We are grateful to Prof.\ T. Ohashi for an insightful question at a
conference which led to the above conductivity estimate.  Support was
provided by NASA contract NAS8-39073, \chandra\ grant GO2-3164X, and the
Smithsonian Institution.

\end{document}